\documentclass[sigconf]{acmart} 
\usepackage[show]{chato-notes}
\usepackage{amsfonts}
\usepackage{graphicx}
\usepackage{fontawesome}
\usepackage{xcolor}

\AtBeginDocument{%
  }

\copyrightyear{2025}
\acmYear{2025}
\setcopyright{rightsretained}
\acmConference[RecSys '25]{Proceedings of the Nineteenth ACM Conference on Recommender Systems}{September 22--26, 2025}{Prague, Czech Republic}
\acmBooktitle{Proceedings of the Nineteenth ACM Conference on Recommender Systems (RecSys '25), September 22--26, 2025, Prague, Czech Republic}\acmDOI{10.1145/3705328.3759317}
\acmISBN{979-8-4007-1364-4/2025/09}

\usepackage{hyperref}
\usepackage{float}
\usepackage{svg}
\usepackage{tabularray}
\UseTblrLibrary{booktabs}

\graphicspath{ {./images/} }
\begin{document}

\title[eSASRec: Enhancing Transformer-based Recommendations in a Modular Fashion]{eSASRec: Enhancing Transformer-based Recommendations in a Modular Fashion}

\author{Daria Tikhonovich}
\affiliation{%
  \institution{MTS}
  \city{Moscow}
  \country{Russia}
}
\email{daria.m.tikhonovich@gmail.com}

\author{Nikita	Zelinskiy}
\affiliation{%
  \institution{MTS}
  \city{Moscow}
  \country{Russia}
}
\email{zelinskiy.nikita.il@gmail.com}

\author{Aleksandr V. Petrov}
\affiliation{%
  \institution{Independent Researcher}
  \city{Glasgow}
  \country{United Kingdom}
}
\email{firexel@gmail.com}

\author{Mayya Spirina}
\affiliation{%
  \institution{MTS}
  \city{Moscow}
  \country{Russia}
}
\email{spirinamay@gmail.com}

\author{Andrei Semenov}
\affiliation{%
  \institution{Yandex}
  \city{Moscow}
  \country{Russia}
}
\email{in48semenov@yandex.ru}
	
\author{Andrey V. Savchenko \orcid{0000-0001-6196-0564}}
\affiliation{%
  \institution{Sber AI Lab}
  \city{Moscow}
  \country{Russia}
}
\email{avsavchenko@hse.ru}

\author{Sergei Kuliev}
\affiliation{%
  \institution{MTS}
  \city{Moscow}
  \country{Russia}
}
\email{kuliev.s.d@gmail.com}

\renewcommand{\shortauthors}{Tikhonovich et al.}

\begin{abstract}
Since their introduction, Transformer-based models, such as SASRec and BERT4Rec, have become common baselines for sequential recommendations, surpassing earlier neural and non-neural methods. 
A number of following publications have shown that the effectiveness of these models can be improved by, for example, slightly updating the architecture of the Transformer layers, using better training objectives, and employing improved loss functions. However, the additivity of these modular improvements has not been systematically benchmarked - this is the gap we aim to close in this paper. 
Through our experiments, we identify a very strong model that uses SASRec's training objective, LiGR Transformer layers, and Sampled Softmax Loss. We call this combination eSASRec (Enhanced SASRec). While we primarily focus on realistic, production-like evaluation, in our preliminarily study we find that common academic benchmarks show eSASRec to be  23\% more effective compared to the most recent state-of-the-art models, such as ActionPiece. In our main production-like benchmark, eSASRec resides on the Pareto frontier in terms of the accuracy–coverage tradeoff (alongside the recent industrial models HSTU and FuXi-$\alpha$). As the modifications compared to the original SASRec are relatively straightforward and no extra features are needed (such as timestamps in HSTU), we believe that eSASRec can be easily integrated into existing recommendation pipelines and can can serve as a strong yet very simple baseline for emerging complicated algorithms. To facilitate this, we provide the open-source implementations for our models and benchmarks in  repository  \faGithubSquare{}~\url{https://github.com/blondered/transformer_benchmark}

\end{abstract}






\maketitle

\section{Introduction and Preliminary Study}
\begin{figure}[tb]
\includegraphics[width=0.8\linewidth]{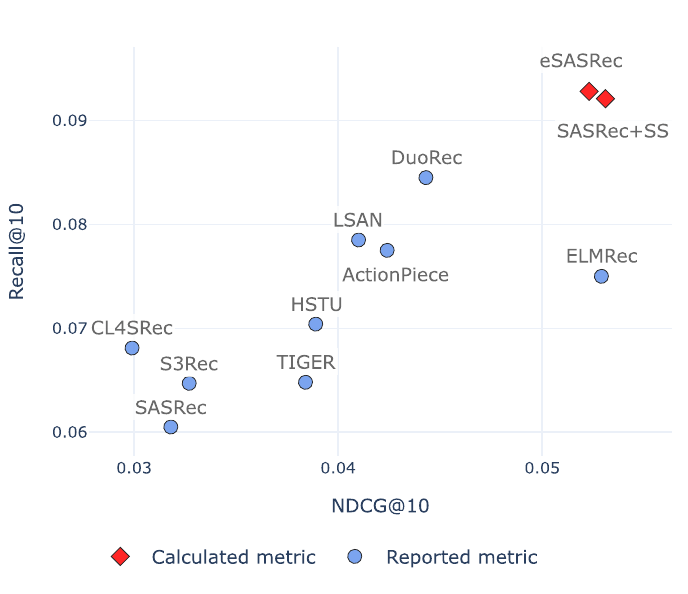}
\caption{The results of our preliminary study on the Amazon Beauty benchmark. Reported metrics are taken from the original papers: S3Rec \cite{s3rec}, LSAN \cite{LSAN}, DuoRec \cite{DuoRec}, CL4SRec \cite{CL4SRec}, TIGER \cite{TIGER}, ActionPiece~\cite{ActionPeace}, ELMRec \cite{ELMRec}. HSTU~\cite{HSTU} and SASRec \cite{SASRec_vanilla} are taken from~\cite{ActionPeace}. }
\label{fig:motivation}
\end{figure}
Sequential recommender systems model the ordered sequence of user-item interactions to predict the next item a user will engage with.
Since the introduction of SASRec~\cite{SASRec_vanilla} in 2018 and BERT4Rec~\cite{BERT4Rec_vanilla} in 2019, these models became very popular in both academic and industrial settings for sequential recommendation. Indeed, these models are routinely used as the baselines in research~\cite{BERT4REC_HF} and their variants are used in production~\cite{koneru2024enhancing}. Despite their popularity, it has been shown that ``vanilla'' versions (i.e. models trained exactly as described in their respective papers) yield sub-optimal effectiveness, which can be improved by employing better training objectives, loss functions and negative sampling strategies~\cite{SASRec_plus, gSASRec, pinner, rss}.

Unfortunately, the additive effect of these modular enhancements has not been systematically studied. In the absence of clear guidance on which combinations to use, most researchers default to standard model configurations \cite{s3rec, ActionPeace, TIGER, ELMRec}. Consequently, improvements over these baselines are often reported as state-of-the-art-claims that, we argue, are frequently overstated. To demonstrate this, we conducted a \emph{preliminary} study comparing enhanced versions of SASRec~\cite{SASRec_vanilla} on the Amazon reviews and MovieLens datasets using the evaluation setup from~\cite{s3rec}. This setup - while not following best practices such as global temporal splitting~\cite{flaws} and thus not used in our main experiments - serves as one of the most common academic benchmarks, enabling direct comparison with previously reported results\footnote{For details of the setup, we refer to the original paper~\cite{s3rec}.}.

Figure~\ref{fig:motivation} presents a subset of our preliminary study results (see Appendix for complete results on five common datasets). The figure compares the reported effectiveness (measured by NDCG@10 and Recall@10) of several recent models with two enhanced versions of SASRec: SASRec+SS (SASRec with Sampled Softmax loss~\cite{gSASRec}) and our proposed eSASRec (SASRec with both Sampled Softmax loss and LiGR~\cite{LiGR} Transformer layers). Both variants outperform most recent models on this dataset. For instance, eSASRec improves NDCG@10 by 23\% over ActionPiece (0.0523 vs. 0.0424). They also substantially outperform vanilla SASRec, confirming the impact of proposed modular improvements.

Our preliminary study motivates us to  evaluate the effectiveness of various modular enhancements to Transformer-based recommender models in production-like scenario with time-based validation strategy and beyond-accuracy (BAc) metrics taken into account. Through a systematic evaluation of multiple possible modifications (including loss functions, training objectives, Transformer layer configurations and negative sampling strategies), we find eSASRec to perform competitively with advanced production models such as HSTU~\cite{HSTU} and FuXi-$\alpha$~\cite{FuXi}. As eSASRec's combination of enhancements has been shown to work on a broad range of datasets, we recommend to use it as a baseline instead of (or alongside with) vanilla versions of Transformer-based models. 

\textbf{In summary, our main contributions are twofold:}
\begin{itemize}
\item We present a realistic evaluation of multiple enhancements for Transformer-based models, including those that have never been tested on public data (e.g. Dense All Action training objective \cite{pinner}, LiGR Transformer layers  \cite{LiGR}).

\item We demonstrate that eSASRec, which enhances original SASRec architecture with LiGR layers and Sampled Softmax loss, is a strong baseline for Transformer-based recommendations.
\end{itemize}

\section{Related Work}

\textit{Transformer-based Models.}
Two main models in sequential recommendation, widely used as standard baselines in the literature, are SASRec \cite{SASRec_vanilla}  that applies a causal self-attentive Transformer decoder with binary cross-entropy (BCE) loss, and BERT4Rec \cite{BERT4Rec_vanilla, BERT4REC_HF}, which uses a bidirectional Transformer encoder with masking-based training objectives. Recent papers demonstrate various enhancements to classical transformer-based recommendation models \cite{CL4SRec, s3rec, TIGER, ActionPeace, LSAN}. 
In particular, \cite{SASRec_plus, SASRec_Dot_MoL, FuXi, effective_sample_softmax} replace softmax cross-entropy over the full item catalog, as used in original BERT4Rec~\cite{BERT4Rec_vanilla}, with SS loss. \cite{gSASRec} introduces a generalized binary cross-entropy (gBCE) loss in place of the BCE loss from SASRec~\cite{SASRec_vanilla}. 
These loss modifications are often paired with negative sampling strategies - ranging from uniform and in-batch (favoring popular items) to more advanced methods. They include logQ correction and other bias-adjusted variants~\cite{ mixed_negative_samplimg, sampling_bias_correction}, hybrid schemes with tunable ratios of uniform and in-batch negatives~\cite{pinner}, and supervised positive sampling~\cite{DuoRec}. 
The authors of~\cite{pinner} revisit the training objective by redefining targets beyond SASRec’s shifted sequence and BERT4Rec’s masking. They propose Next Action (predicting a user's last event), All Action - predicting all events in a holdout period, and Dense All Action (DenseAA) - sliding-window prediction. 
Recent studies introduced recommendation task-specific transformer architectures, which demonstrated superior quality: FuXi-$\alpha$ \cite{FuXi}, LiGR Transformer \cite{LiGR}, Hierarchical Sequential Transduction Unit Encoder \cite{HSTU}, etc. One could also find modifications in positional encoding \cite{pos_encoding}, tokenization techniques \cite{s3rec, DuoRec}, features and context integration \cite{features}, data augmentation \cite{CBiT, tada, da_survey}, different similarity functions~\cite{Rendal, cosine}, etc. The wide variety of such enhancements motivated us to categorize them systematically and adopt a modular approach in our experiments. 

\textit{Splitting strategy.} 
Leave-one-out (LOO) is the dominant strategy for Transformer-based recommendations \cite{BERT4REC_HF, SASRec_plus, SASRec_Dot_MoL, s3rec, HSTU, TIGER}. Still, it suffers from temporal leakage - test interactions may overlap with training data~\cite{flaws}. Several works advocate for time-based splits \cite{flaws, leak}, which have been partly adopted by studies like \cite{temporal_context, bpr2, large_item_catalog}. As LOO is not realistic and does not follow recent recommendations, we do not use it in our main set of experiments, and employ global temporal split instead.

\section{Methodology}

\label{subsec:modular_enhancements}

\textit{Modular approach.} In this paper, we adopt a modular view on Transformer-based models from \cite{tr4rec} and \cite{pinner}, bringing each architecture to interchangeable modules. For example, while SASRec~\cite{SASRec_vanilla} applies a shifted sequence training objective and BERT4Rec~\cite{BERT4Rec_vanilla} applies item masking, both models can be trained with multiple loss variants \cite{gSASRec} or negative sampling strategies \cite{pinner}. Sometimes, the Transformer layers' architecture is also updated \cite{LiGR, BERT4REC_HF, tr4rec}.

We examine multiple variants for each of its modular components. In contrast to \cite{tr4rec} (experimenting with available HuggingFace architectures), in our study we focus on approaches with successful reported results in industrial A/B tests: gSASRec~\cite{gSASRec, sasrecbias}, HSTU~\cite{HSTU}, FuXi-$\alpha$\cite{FuXi}, PinnerFormer \cite{pinner}, and LiGR~\cite{LiGR}. Below, we briefly outline the selected modular enhancements from these papers.

\begin{itemize}
\item \textbf{Training objective}.
We include standard objectives like Shifted Sequence from SASRec \cite{SASRec_vanilla} and MLM from BERT4Rec \cite{BERT4Rec_vanilla}), along with less typical variants: Next Action, All Action, and DenseAA from \cite{pinner}, which predicts the last interaction, all interactions in a holdout period, or dense interactions within a sliding window, respectively. 

\item \textbf{Transformer layers architecture}.
We evaluate the Post-LN from SASRec \cite{SASRec_vanilla} and LiGR \cite{LiGR} architectures.

\item \textbf{Loss options.}
We evaluate two losses based on negative sampling: generalized binary cross-entropy (gBCE) introduced in \cite{gSASRec} and used by \cite{sasrecbias}, and a widely adopted SS \cite{ SASRec_Dot_MoL, FuXi, pinner}.

\item \textbf{Negative sampling strategies.} We include the common uniform negatives sampling  \cite{SASRec_vanilla, BERT4Rec_vanilla} as well as in-batch sampling and a hybrid method, combining uniform and in-batch sampling with adjustable ratios  \cite{pinner, mixed_negative_samplimg}. We optionally apply logQ correction \cite{logQ_old, sampling_bias_correction}. 

\end{itemize}



\textit{eSASRec as the winning combination.} Since our experiments reveal the best-performing modular architecture based on SASRec's Shifted Sequence training objective~\cite{SASRec_vanilla}, we define it as Enhanced SASRec (eSASRec). We now briefly describe the key changes from the original SASRec model.

First, we replace the original SASRec Transformer layers with the LiGR architecture proposed in \cite{LiGR}. 
LiGR Transformer blocks imply pre-norm and each multi-head attention
and the feed-forward layer is gated with a linear projection and sigmoid activation:
\begin{equation}
h^{j+1} = h^j + F(h^j) \times \sigma(h^jW),
\end{equation}
where $F$ represents a multi-head attention or a feed-forward layer. While the original paper \cite{LiGR} did not provide more details on implementation, we use the SwiGLU \cite{glu} activation function in feed-forward layers. 


We adopt the Sampled Softmax loss~\cite{SASRec_plus, SASRec_Dot_MoL}, which showed results comparable to gBCE~\cite{gSASRec}, but offers broader adoption~\cite{FuXi, HSTU, SASRec_Dot_MoL} and consistent Pareto efficiency. Third, to enhance BAc performance, we optionally apply mixed negative sampling (MN). While the original SASRec uses negatives sampled uniformly from the full item set, MN includes a proportion of in-batch items~\cite{pinner}, following prior industrial practice~\cite{HSTU, pinner}. This approach improves the balance between accuracy and personalization. Our final eSASRec architecture combines the Shifted Sequence objective from SASRec with LiGR layers and Sampled Softmax loss (SASRec+LiGR+SS), with optional MN sampling.

\textit{Production-like Realistic Benchmark.} Transformer-based models are now widely used in production systems and existing frameworks~\cite{pinner, FuXi, HSTU, CL4SRec, LiGR, tr4rec}, yet offline evaluation remains limited~\cite{flaws}. To enable realistic model evaluation, we adopt a time-based validation strategy~\cite{flaws}, which mitigates data leakage in LOO setups~\cite{temporal_context, leak, flaws, val} and is gaining adoption~\cite{bpr2}. We report both accuracy and BAc metrics, since BAc quality was shown to improve user engagement in multiple production settings \cite{field_23_nastya, pinner, sasrecbias}.

\section{Experimental Evaluation}
\subsection{Setup}


Our experiments aim to address the following research questions:



    
\textbf{RQ1.} Which modular enhancements improve Transformer models performance?

\textbf{RQ2.} Which modular combination should be used as the effective baseline in experiments?
    


\begin{table}[tb]
    \centering
    \caption{Dataset Summary Statistics}
    \label{tab:dataset_statistics}
    \begin{tabular}{lcccccc}
        \toprule
        \textbf{Dataset} & \textbf{Users} & \textbf{Items} & \textbf{Interactions} & \textbf{Avg. Len.} &  \\
        \midrule
        ML-20M  & 138493 & 26744 & 20,000,263 & 144.4  \\
        Kion  & 606743 & 10266 & 5,109,907 & 8.4 \\
       BeerAdvocate  & 22679 & 22264 & 1,498,969 & 66.1 \\
        \bottomrule
    \end{tabular}
\end{table}

\begin{table*}[ht!]
    \centering
    \caption{Performance in realistic setting in terms of NDCG@10 (N@10), Coverage@10 (Cov@10) and Pareto-optimality (Par). }
    \label{tab:prod_performance_datasets}
    \resizebox{0.95\linewidth}{!}{
    \begin{booktabs}{llccccccccc}
      \toprule
       ~ & ~ &    \SetCell[c=3]{c}{{{ML-20M}}} & & & \SetCell[c=3]{c} {{{Kion}}} & & & \SetCell[c=3]{c} {{{BeerAdvocate}}}  \\
       \cmidrule[lr]{3-5} \cmidrule[lr]{6-8} \cmidrule[lr]{9-11}
        \textbf{Modules}& \textbf{Model}& \textbf{N@10}& \textbf{Cov@10}& \textbf{Par}& \textbf{N@10}& \textbf{Cov@10}& \textbf{Par}&
        \textbf{N@10}& \textbf{Cov@10}& \textbf{Par}\\
        \midrule
        Baselines& Popular-7-days& 0.1671 & 0.0061 & & 0.142 & 0.0035 & & 0.0575 & 0.0013 & \\

        & SASRec Vanilla, BCE 1 neg& 0.1192 & 0.0829 & & 0.1389 & 0.113 & & 0.0222 & 0.0078 & \\
        \midrule
        SOTA & HSTU & 0.1872 & 0.0535 & {\Large\textcolor{green!60!black}{\checkmark}} & \textbf{0.1694} & 0.2110 &{\Large\textcolor{green!60!black}{\checkmark}}  & 0.0473 & \textbf{0.1942}& {\Large\textcolor{green!60!black}{\checkmark}}  \\
        
        models& FuXi-$\alpha$& \textbf{0.1883} & 0.0513 &{\Large\textcolor{green!60!black}{\checkmark}}  & 0.1647 & 0.2319 & & 0.0472 & 0.1904& \\
        \midrule
        Layers 
        & SASRec+SS& 0.1527 & 0.0913 & {\Large\textcolor{green!60!black}{\checkmark}}  & 0.1497 & 0.0785 & & 0.0381 & 0.0079&\\
        
        \& Losses & SASRec+LiGR+SS (eSASRec) & 0.1563 &   0.0889 & {\Large\textcolor{green!60!black}{\checkmark}} & 0.1657 & 0.3003 & {\Large\textcolor{green!60!black}{\checkmark}}   & 0.0650 & 0.0771& {\Large\textcolor{green!60!black}{\checkmark}}  \\
        
        & SASRec+LiGR+gBCE-0.75& 0.1479 & 0.1049 & {\Large\textcolor{green!60!black}{\checkmark}} & 0.1685 & 0.2381 & {\Large\textcolor{green!60!black}{\checkmark}} & 0.0629 & 0.0737 &  \\
        \midrule
        
        & SASRec+LiGR+SS+Mixed-0.6& 0.1248 & \textbf{0.1076} & {\Large\textcolor{green!60!black}{\checkmark}}  & 0.1475 & 0.4049 & {\Large\textcolor{green!60!black}{\checkmark}}  & 0.0606 & 0.0746 & \\

        Negatives& SASRec+LiGR+SS+Mixed-0.6-LogQ& 0.1358 & 0.0589 & & 0.1621 & 0.1218 &  &
        \textbf{0.0690} & 0.0271& {\Large\textcolor{green!60!black}{\checkmark}} \\

        & SASRec+LiGR+SS+InBatch& 0.0020 & 0.0245 & & 0.1283 & \textbf{0.6055} &{\Large\textcolor{green!60!black}{\checkmark}}  & 0.0310 & 0.0489&  \\
        \midrule
        
         & DenseAA+LiGR+SS& 0.1668 & 0.0657 & {\Large\textcolor{green!60!black}{\checkmark}}  & 0.1638 & 0.1906 & & 0.0635 & 0.0354 & \\
        Training& AllAction-Causal+LiGR+SS & 0.1481 & 0.0063 & & 0.1536 & 0.1473 & & 0.0300 & 0.0020 & \\
        
        objectives& NextAction-Causal+LiGR+SS& 0.1373 & 0.0329 & & 0.1636 & 0.1431 & & 0.0423 & 0.0053 & \\
        
        & BERT4Rec+LiGR+SS& 0.1364 & 0.0808 & & 0.1598 & 0.1324 & & 0.0561 & 0.0353&  \\
        \midrule
        
        Mixing& DenseAA+LiGR+gBCE-0.75& 0.1762 & 0.0580 & {\Large\textcolor{green!60!black}{\checkmark}} & 0.1669 & 0.1576 &  & 0.0576 & 0.0283& \\
        \bottomrule
    \end{booktabs}
}
\end{table*}

\textit{Splitting strategy.} For generating a time-based test set, we select a time window in days from options \{14, 30, 60\} closer to reflecting 5\% of each dataset's interaction count. We use all interactions within the selected time window for testing and those prior for training, filtering out test interactions involving users or items absent from the training set. For hyper-parameters tuning and model early stopping we generate validation fold from training set using the LOO strategy since our experiments showed time-based splitting for validation fold to hurt model performance on test set due to more interactions being dropped from training. We also confirmed that LOO strategy for parameters tuning had negligible impact on final parameter values within one model class.

\textit{Datasets.}
Following prior work, we use ML-20M dataset due to its broad adoption \cite{HSTU, SASRec_Dot_MoL, SASRec_plus, FuXi}. We also use Kion~\cite{kion} dataset which is based on implicit feedback instead of ratings and BeerAdvocate \cite{beer} dataset due to sufficient average sequence length which greatly exceeds the most common Amazon Review datasets \cite{ActionPeace, TIGER, ELMRec, DuoRec, LSAN, s3rec}. Following \cite{flaws}, we apply 2-core filtering for users and 5-core filtering for items. The test window size for splitting is 60 days for Ml-20m and BeerAdvocate datasets and 14 days for Kion dataset.

\hyphenation{Trans-former}
\textit{Baselines.} As the state-of-the-art baselines we use the recent production-grade Transformer-based models HSTU \cite{HSTU} and FuXi-$\alpha$ \cite{FuXi}, reported successful in both in real-life A/B tests and on common academic benchmarks. As an additional baseline, we include the commonly used \cite{HSTU, FuXi, SASRec_Dot_MoL} SASRec variant with the SS loss (SASRec+SS) \cite{sasrecbias} as one of the possible modifications for SASRec model. For completeness, we we also use the original version of SASRec trained with BCE loss and 1 sampled negative \cite{SASRec_vanilla} and the \textit{Popular} baseline, which recommends the most popular items from the last 7 days of training interactions.

\textit{Implementation Details.}
We use the RecTools\footnote{\url{https://github.com/MobileTeleSystems/RecTools}. The model implementations in this framework are trustworthy, as they have been validated against reported results~\cite{tikhonovich2023rectools}.} PyTorch implementations of SASRec and BERT4Rec backbone models with standard loss functions available out of the box (SS, gBCE, BCE) and implement all required modular modifications on top of them. For HSTU \cite{HSTU} and FuXi-$\alpha$ \cite{FuXi} algorithms we use the original authors implementations and model training code. For all datasets we first tune hyper-parameters for SASRec model trained with sampled softmax loss (SASRec+SS), which was shown to be effective in prior works \cite{SASRec_plus, SASRec_Dot_MoL, HSTU}. We used the following grid: hidden size options: 50, 64, 128, 256, number of self-attention blocks options: 1, 2, 4, number of attention heads options: 1, 2, 4, 8, dropout rate options: 0.1, 0.2, 0.3, 0.5. We select maximum sequence length, maximum number of epochs and patience based on the dataset. These settings and selected parameters for ML-20m are consistent with the
previous papers \cite{HSTU, SASRec_plus, SASRec_Dot_MoL}. Final parameters for each dataset can be found in \autoref{tab:model_hyperparameters}.
Following \cite{HSTU, FuXi}, we keep common hyper-parameters fixed for each new model modification, as well as for HSTU and FuXi-$\alpha$, but tune modification-specific parameters (like time-window for AllAction training objective). For training we use batch size of 128 and Adam optimizer with the learning rate 1e-3. 
The best model checkpoint is selected based on the validation loss instead of Recall@10, since we are also interested in BAc quality.

\textit{Metrics.} We report accuracy with NDCG@10, measure global diversity of recommendations across users with Coverage@10 \cite{coverage_2010, pinner, curse_low}, and evaluate performance in terms of Pareto-optimality.

\subsection{Results} 
\begin{figure}[tb!]
\includegraphics[width=0.8\linewidth]{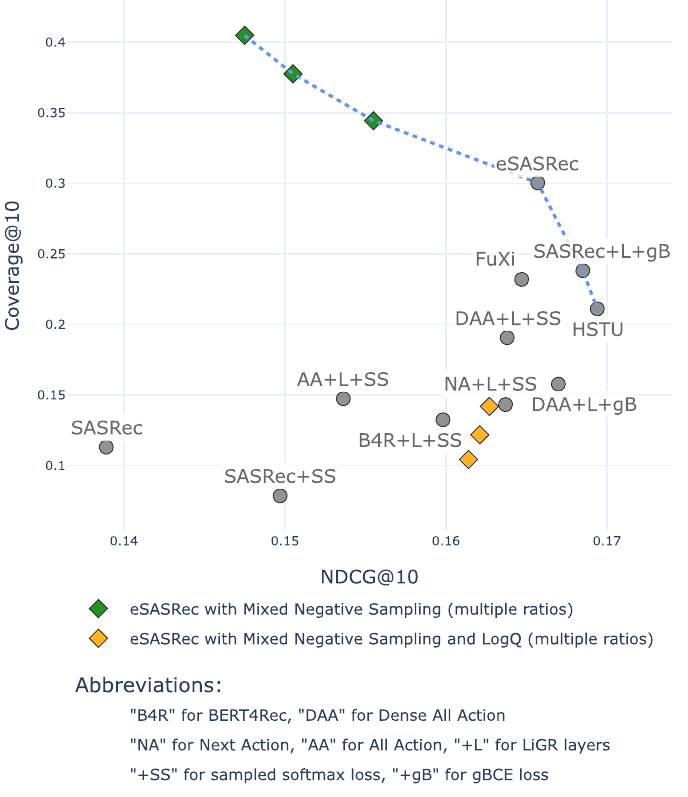}
\caption{NDCG/Coverage Pareto front on Kion Dataset}
\label{fig:pareto}
\end{figure}

\begin{table}
    \centering
    \caption{Model Hyper-parameters}
    \label{tab:model_hyperparameters}
    \begin{tabular}{lccc}
        \toprule
        \textbf{Hyper-parameter} & \textbf{ML-20M} & \textbf{Kion} & \textbf{BeerAdvocate}\\
        \midrule
        emb\_dim & 256 & 256 & 256\\
        n\_blocks & 4 & 2 & 4\\
        n\_heads & 8 & 2 & 2\\
        dropout\_rate & 0.2 & 0.1 & 0.3\\
        ff\_emb\_mult (LiGR only) & 4 & 4 & 4\\
        sequence\_max\_len & 200 & 50 & 100\\
        n\_negatives & 256 & 256 & 256\\
        max\_epochs & 100 & 100 & 100\\
        patience & 50 & 50 & 10\\
        learning\_rate & 0.001 & 0.001 & 0.001\\
        \bottomrule
    \end{tabular}
\end{table}

Table \ref{tab:prod_performance_datasets} presents our results for evaluated modifications and reveals the Pareto-optimality of each model in terms of NDCG-Coverage trade-off. For brevity we do not include results for evaluated combinations with decreased performance. 
Due to presence of counterbalancing metrics and following prior work, we do not provide explicit statistical tests results \cite{HSTU, SASRec_Dot_MoL}. Concerning RQ1, we see the following effect from specific modular enhancements:


\textit{Modifications of Training Objective.}
Regarding training objective modifications, no major improvements are observed over the SASRec's~\cite{SASRec_vanilla} Shifted Sequence objective as no other variants retain on the Pareto frontier for multiple datasets. DenseAA training objective shows comparable performance on ML-20M dataset while other variants consistently underperform. 

\textit{Loss Functions.}
The \textit{SASRec+SS} enhancement outperforms the original SASRec (with BCE loss and one negative) on both NDCG and Coverage, demonstrating overall higher performance. Switching from SS loss to gBCE with a temperature of 0.75~\cite{gSASRec} allows to control the BAc tradeoff on 2 out of 3 datasets but slightly moves the model from to the Pareto-front on the BeerAdvocate dataset.

\textit{Transformer layers: Vanilla SASRec vs. LiGR.}
\looseness -1 When comparing the SASRec+SS \cite{SASRec_plus} with vanilla Transformer layers to the added LiGR~\cite{LiGR} enhancement (SASRec+LiGR+SS), the results on ML-20M are similar. At the same time, on Kion and BeerAdvocate we see a major increase in NDCG and Coverage: e.g NDCG@10 increases by 9\% and Coverage increases by 280\% on the Kion dataset. Hence, we suggest LiGR layers to be an effective enhancement for Transformer baselines compared to the original Transformer architecture.

\textit{Negative Sampling Strategies.}
Most of the time introducing or expanding a share of in-batch negatives compared to original uniform sampling decreases NDCG while increasing Coverage. More detailed results on these experiments are presented in our repository. We conclude that incorporating MN sampling and tuning the in-batch negative ratio and presence or absence of LogQ correction may have considerable impact on the final metrics trade-off. 

\textit{RQ2. Effective combination.} Only two models remain on the Pareto-front for all  datasets: the SASRec+LiGR+SS combination and HSTU \cite{HSTU}. We find that FuXi-$\alpha$ \cite{FuXi} and SASRec+LiGR+gBCE have a very slight drop of performance on some datasets. Therefore, we propose SASRec+LiGR+SS combination as the effective baseline for academic and industrial experiments and we call this combination eSASRec (Enhanced SASRec). The example of Pareto front for Kion dataset is presented on \autoref{fig:pareto}. Since the Pareto front always includes some variants of MN sampling, we propose this enhancement as optional for any model.

\section{Conclusion and future work}

In this paper, we evaluated multiple modular enhancements for Transformer-based models in search for baseline architecture with the strongest performance. The winning eSASRec model combined SASRec's training objective with LiGR Transformer layers and Sampled Softmax loss. This combination showed 23\% improvements over the recent State-of-the-Art approaches in our preliminary experiments on common academic benchmark and demonstrated performance comparable with the up-to-date industrial model HSTU in realistic evaluation settings. 

We provide model implementations and evaluation code along with processed datasets and data splits to foster transparency and facilitate future work. More experiments on even larger datasets and with the use of different from our work modular modifications could help to establish architectures with consistently strong performance. Moreover, while Coverage metric is widely-used, it is important to add other relevant BAc metrics, such as Diversity, Serendipity, and Novelty, into future benchmarks.

Our experiments under realistic settings did not confirm the performance gain of eSASRec over HSTU shown in common academic benchmarks on Amazon review datasets, which are widely used for state-of-the-art claims (our preliminary study, \autoref{fig:motivation}). This emphasizes the need for further research in the field of rigorous evaluation of new algorithms. We hope that our findings reinforce the importance of strong baselines and comprehensive evaluation in recommender systems.

\section*{Appendix}

Detailed metrics, Recall@10 (R@10) and NDCG@10 (N@10), for five common datasets from our preliminary study  are shown in \autoref{tab:performance_comparison_beauty_sports_toys_transposed} and \autoref{tab:performance_comparison_ml_base_transposed}. Here, all metrics in part ``Our Experiments'' are obtained by training all models (HSTU, FuXi-$\alpha$, SASRec+SS, eSASRec) with identical hyper-parameters within each dataset.


\begin{table}[ht!]
\centering
\caption{Preliminary study results on Amazon datasets.} 
\label{tab:performance_comparison_beauty_sports_toys_transposed}
\resizebox{\linewidth}{!}{
\begin{tabular}{l|c|cc|cc|cc}
\toprule
& & \multicolumn{2}{c|}{\textbf{Beauty}} & \multicolumn{2}{c|}{\textbf{Sports}} & \multicolumn{2}{c}{\textbf{Toys}} \\
\cmidrule(lr){3-4} \cmidrule(lr){5-6} \cmidrule(lr){7-8}
\textbf{Model} & \textbf{Source} & \textbf{R@10} & \textbf{N@10} & \textbf{R@10} & \textbf{N@10} & \textbf{R@10} & \textbf{N@10} \\

\midrule
\multicolumn{8}{c}{Reported in Literature} \\
\midrule
SASRec & \cite{s3rec}& 0.061 & 0.032 & 0.035 & 0.019 & 0.068 & 0.037 \\
\midrule
S3Rec  & \cite{s3rec} & 0.065 & 0.033 & 0.039 & 0.020 & 0.070 & 0.038 \\
\midrule
LSAN  & \cite{LSAN} & 0.079 & 0.041 & 0.048 & 0.026 & 0.071 & 0.037 \\
\midrule
DuoRec & \cite{DuoRec} & 0.085 & 0.044 & 0.050 & 0.026 & - & - \\
\midrule
CL4SRec & \cite{CL4SRec} & 0.068 & 0.030 & 0.039 & 0.017 & - & - \\
\midrule
TIGER & \cite{TIGER} & 0.065 & 0.038 & 0.040 & 0.023 & 0.071 & 0.043 \\
\midrule
ActionPiece & \cite{ActionPeace} & 0.078 & 0.042 & 0.050 & 0.026 & - & - \\
\midrule
HSTU & \cite{ActionPeace} & 0.070 & 0.039 & 0.041 & 0.022 & - & - \\
\midrule
\multicolumn{8}{c}{Our Experiments} \\
\midrule
HSTU & & 0.079 & 0.044 & 0.044 & 0.024 & 0.078 & 0.045 \\
\midrule
FuXi-$\alpha$ & & 0.081 & 0.045 & 0.044 & 0.024 & 0.077 & 0.044 \\
\midrule
SASRec+SS & & 0.092 & \textbf{0.053} & \textbf{0.057} & 0.031 & \textbf{0.097} & \textbf{0.058} \\
\midrule
eSASRec & & \textbf{0.093} & 0.052 & 0.056 & \textbf{0.032} & 0.094 & 0.054 \\
\bottomrule
\end{tabular}}
\end{table} 

\begin{table}[ht!]
\centering
\caption{Preliminary study results on MovieLens datasets. ML-20M-Large stands for 8-layers-scaled model results \cite{HSTU, FuXi}.}
\label{tab:performance_comparison_ml_base_transposed}
\resizebox{\linewidth}{!}{\begin{tabular}{l|c|cc|cc|cc}
\toprule
&& \multicolumn{2}{c|}{\textbf{ML-1M}} & \multicolumn{2}{c|}{\textbf{ML-20M}} & \multicolumn{2}{c}{\textbf{ML-20M-Large}} \\
\cmidrule(lr){3-4} \cmidrule(lr){5-7} \cmidrule(lr){7-8}
\textbf{Model} & \textbf{Source} & \textbf{R@10} & \textbf{N@10} & \textbf{R@10} & \textbf{N@10} & \textbf{R@10} & \textbf{N@10} \\
\midrule
\multicolumn{8}{c}{Reported in Literature} \\
\midrule
SASRec MoL &\cite{SASRec_Dot_MoL}& 0.308 & - & 0.311 & - & - & - \\
\midrule
HSTU & \cite{HSTU} & 0.310 & 0.172 & 0.325 & 0.188 & \textbf{0.357} & \textbf{0.211} \\
\midrule
FuXi-$\alpha$  & \cite{FuXi}& - & - & 0.335 & 0.195 & 0.353 & 0.209 \\
\midrule
\multicolumn{8}{c}{Our Experiments} \\
\midrule
HSTU&& 0.304 & 0.172 & 0.343 & 0.203 & 0.346 & 0.204 \\
\midrule
FuXi-$\alpha$ && \textbf{0.316} & \textbf{0.181} & \textbf{0.355} & \textbf{0.212} & 0.347 & 0.204 \\
\midrule
SASRec+SS && 0.293 & 0.168 & 0.313 & 0.183 & 0.047 & 0.023 \\
\midrule
eSASRec && 0.313 & 0.177 & 0.329 & 0.197 & 0.346 & \textbf{0.211} \\
\bottomrule
\end{tabular}}
\end{table}

\bibliographystyle{ACM-Reference-Format}
\bibliography{refs}

\end{document}